APPLIED ACOUSTICS

# Laser streaming: Turning a laser beam into a flow of liquid

Yanan Wang,[1,3]* Qiuhui Zhang,[2,3]* Zhuan Zhu,[3] Feng Lin,[1,3] Jiangdong Deng,[4] Geng Ku,[5] Suchuan Dong,[6] Shuo Song,[3] Md Kamrul Alam,[7] Dong Liu,[8] Zhiming Wang,[1]† Jiming Bao[1,3,7]†

Transforming a laser beam into a mass flow has been a challenge both scientifically and technologically. We report the discovery of a new optofluidic principle and demonstrate the generation of a steady-state water flow by a pulsed laser beam through a glass window. To generate a flow or stream in the same path as the refracted laser beam in pure water from an arbitrary spot on the window, we first fill a glass cuvette with an aqueous solution of Au nanoparticles. A flow will emerge from the focused laser spot on the window after the laser is turned on for a few to tens of minutes; the flow remains after the colloidal solution is completely replaced by pure water. Microscopically, this transformation is made possible by an underlying plasmonic nanoparticle-decorated cavity, which is self-fabricated on the glass by nanoparticle-assisted laser etching and exhibits size and shape uniquely tailored to the incident beam profile. Hydrophone signals indicate that the flow is driven via acoustic streaming by a long-lasting ultrasound wave that is resonantly generated by the laser and the cavity through the photoacoustic effect. The principle of this light-driven flow via ultrasound, that is, photoacoustic streaming by coupling photoacoustics to acoustic streaming, is general and can be applied to any liquid, opening up new research and applications in optofluidics as well as traditional photoacoustics and acoustic streaming.

## INTRODUCTION

Although the term "ray" has been used to describe sunlight through windows or clouds, it found the perfect match with the invention of laser. The ray of a laser implies not only its high intensity and directionality but also its mechanical impetus because photons carry linear momentum. Is it possible to transfer this unique character of a laser to matter and generate rays of matter such as a liquid flow or stream? Through efficient momentum transfer, the answer is yes. The radiation pressure of a laser beam has been used to deform a liquid surface and even create a stream, but the conditions are difficult to achieve because that liquid would not only need to scatter light strongly but also have near-zero surface tension (1–5). Most liquids do not have these properties; any liquid that has them is actually a mixture of two liquids near the two-phase critical transition for maximum light scattering, and it also requires a liquid-liquid interface to minimize surface tension (1–3). Because of these problems, this momentum transfer technique, no matter if it is a direct transfer from incident photons or an indirect transfer through fluorescent photons (6), has almost no practical applications because common liquids such as water and many organic solvents are homogeneous and do not scatter light at all. Even if all the momentum of a laser beam is transferred to a liquid, it is still too weak to create a stream strong enough to drive a microfluidic device (7, 8).

Many other related techniques have also been developed to manipulate matter with light. Radiometric force or photophoresis allows matter to be transported with laser beams, but it is only limited to low-density materials in gas media (9). Colloidal particles have been used to control fluid motion through optical tweezing (10–12). Cavitation can be used to mix fluids (13, 14) or generate droplets (15), whereas the transportation of droplets can be realized using either the optothermocapillary (16–18) or chromocapillary (19, 20) effect. Note that the driving force behind many optothermocapillary or chromocapillary effects is the well-known Marangoni effect, that is, temperature-dependent surface tension of fluids. Thus, these techniques can only be used to manipulate droplets with photosensitive surfactants (16–20). Here, we report a new optofluidic principle that can be used to generate high-speed flows inside any liquids without any chemical additives or apparent visible moving mechanical parts. The speed, direction, and size of the flow can be controlled by the laser. No prefabricated glass substrate is needed as long as the liquid is optically accessible through a glass window.

## RESULTS AND DISCUSSION

Figure 1 (A and B) shows the experimental setup. When the cuvette is filled only with deionized (DI) water, the laser passes through the cuvette without any attenuation. Significant attenuation is observed when 50-nm gold nanoparticles are added to the water at a concentration of 0.03 mg/ml ($2.4 \times 10^{10}$/ml Au nanoparticle concentration and $1.55 \times 10^{-6}$ Au volume fraction). This is because these nanoparticles exhibit a strong localized surface plasmon resonance around 525 nm, which allows them to effectively absorb the incident laser. Surprisingly, water near the laser spot on the cuvette seemed to be pushed away from the cuvette wall by the laser after a few minutes of illumination, and after 10 more minutes, an elongated stream emerged. Figure 1C shows a snapshot of the flow with the laser at normal incidence. Here, polymer microspheres were added to the water to visualize the flow and focusing was slightly adjusted to maximize the flow speed. To generate a flow at an angle relative to the cuvette surface, we can tilt the cuvette, as shown in Fig. 1B, and focus the laser on a new spot on the cuvette. Figure 1 (D to F) shows representative

[1]Institute of Fundamental and Frontier Sciences, University of Electronic Science and Technology of China, Chengdu, Sichuan 610054, China. [2]Department of Electrical Information Engineering, Henan University of Engineering, Xinzheng, Henan 451191, China. [3]Department of Electrical and Computer Engineering, University of Houston, Houston, TX 77204, USA. [4]Center for Nanoscale Systems, Harvard University, Cambridge, MA 02138, USA. [5]Department of Mechanical Engineering, University of Kansas, Lawrence, KS 66045, USA. [6]Department of Mathematics, Purdue University, West Lafayette, IN 47907, USA. [7]Materials Science and Engineering, University of Houston, Houston, TX 77204, USA. [8]Department of Mechanical Engineering, University of Houston, Houston, TX 77204, USA.
*These authors contributed equally to this work.
†Corresponding author. Email: jbao@uh.edu (J.B.); zhmwang@uestc.edu.cn (Z.W.)

flows at incident angles of 10°, 20°, and 30°, respectively. In all these cases, the flows appear as liquid analogs of laser beams and move in the same directions of the refracted beams as if they are directly driven by photons of laser beams. We call this phenomenon laser streaming.

To better visualize the flow and the circulation of surrounding water, we used a cylindrical lens to focus a 633-nm HeNe laser to generate a light sheet on the focal plane of the video camera. Figure 2 (A and B) shows the same flow in Fig. 1C near the back and front walls of the cuvette. An impinging flow pattern can be observed at the back wall. Figure 2 (B to H) shows that as laser powers decreased, the speed of flow also decreased until the power dropped to 60 mW when there is no flow at all. It is important to note that Au nanoparticles are necessary to create a flow, but once a flow is formed, it stays the same after the replacement of the Au nanoparticle suspension with DI water in the same cuvette. We can conclude from these observations that the flow is not driven by the momentum of the laser beam, and Au nanoparticles must have made permanent changes to the front cuvette glass wall when the laser was on.

The changes can be confirmed by simple visual inspection possible with the naked eye. Optical and electron microscopies reveal that volcano crater–like cavities have been created on the inner surface of the front cuvette glass by the focused laser and Au nanoparticles. Scanning electron microscopy (SEM) images in Fig. 3 (A and B) show that a cavity has a size similar to that of the laser spot and has a very corrugated surface. High-resolution SEM in Fig. 3 (C to D) further reveals that the surface of the cavity is decorated with high-density Au nanoparticles. To obtain better depth and surface topology of the cavity, we imaged cavities with an optical profilometer. Figure 3 (E to H) shows depth profiles of four cavities made by the laser at incident angles of 0°, 10°, 20°, and 30°, respectively. The correlation between laser incidence angle and cavity profile is clear: The cavity becomes more and more eccentric as the incident angle increases.

Now, the mystery begins to unfold: The Au particle–decorated cavity must be responsible for the laser streaming. On one hand, it is well established that when a pulsed laser is incident on plasmonic nanoparticles, ultrasonic waves will be generated via the photoacoustic effect due to rapid heating and cooling accompanied by rapid volume expansion and contraction of the nanoparticles and their surrounding media (21–27). The spectrum of such laser-induced ultrasound is broad, with its central frequency determined mainly by the temporal profile of input laser pulses (26, 27). On the other hand, it was first observed centuries ago by Faraday and Dvorak but is still under active research that ultrasound can generate a steady-state flow, a phenomenon called acoustic streaming (28–31). Nevertheless, a simple combination of these two effects cannot guarantee a flow. For instance, laser-induced ultrasound from the suspended Au nanoparticles cannot generate a flow in the cuvette. To identify the ultrasound mechanism for the laser streaming, we used a hydrophone (V312-SU-F0.46-IN-PTF from Olympus, 10-MHz bandwidth) to analyze the ultrasounds from nanoparticles and the cavity. Figure 4A shows ultrasound traces from the Au nanoparticle solution before and after a flow is created. The difference is clear: When there is no flow, ultrasound is dominated by two short pulses, but the ultrasound with a flow is long-lived and lasts several hundred nanoseconds. Furthermore, the ultrasound with a flow exhibits periodic modulation, which can be observed by zoomed-in traces and fast Fourier transform (FFT) analysis in Fig. 4B. The central frequency of ~1.1 MHz is consistent with 150 ns of laser pulse width (26, 27).

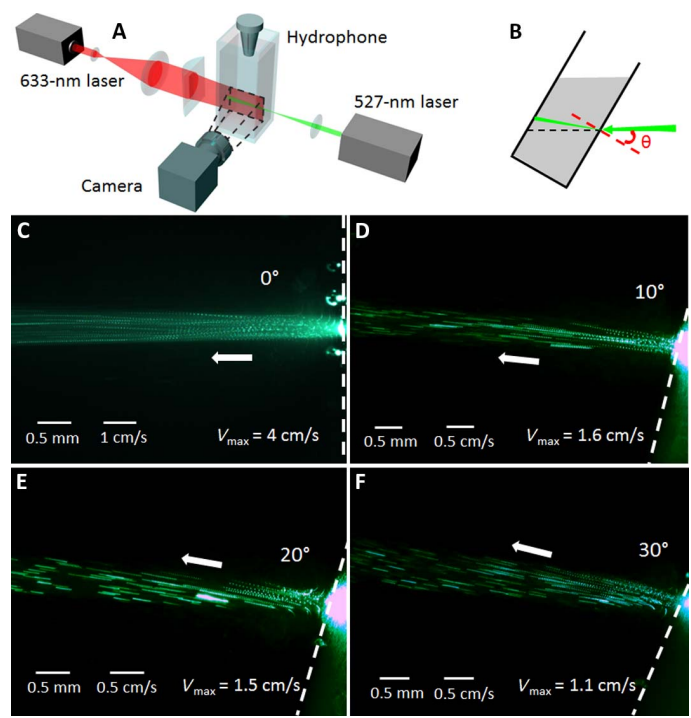

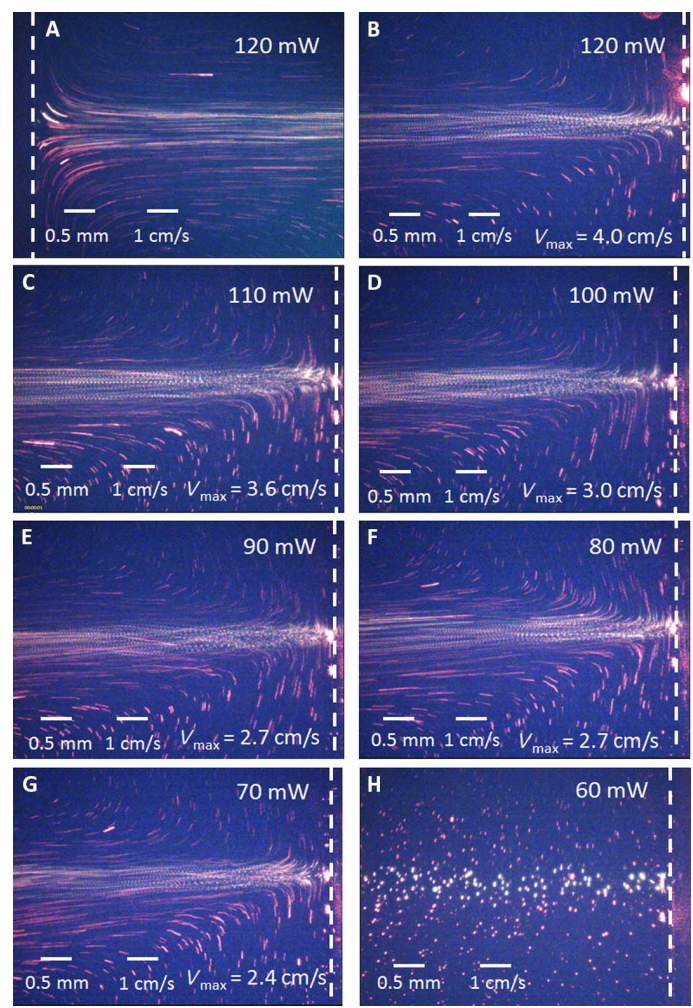

**Fig. 1. Experimental setup and streams at different incident angles of lasers.** (**A**) Schematic. A video camera with variable exposure time is used to capture the motion of streams. The thickness of the cuvette wall is 1 mm. (**B**) The cuvette is tilted to vary the incident angle. (**C** to **F**) Optical images of streams at incident angles of (C) 0°, (D) 10°, (E) 20°, and (F) 30°. White dashed line indicates the cuvette surface, and white block arrows indicate the directions of flows and laser propagation. The angle in the water becomes smaller because of refraction. Laser power is 120 mW. The exposure time for each shot is 100 ms. The trajectory lengths of polymer microspheres are proportional to the flow speed at that point and can be calculated from speed scale bars.

**Fig. 2. Flows and flow patterns at normal incidence under decreasing laser powers.** A 633-nm HeNe laser was used to illuminate tracing microspheres, and a long-pass filter was used to block 527-nm light. The optical path length of the cuvette is 1 cm. The exposure time is 50 ms. White dashed lines indicate cuvette surfaces, and (**A**) is downstream and (**B**) is upstream of the flow under 120-mW laser. (**C**) to (**H**) are upstream flows under different laser powers as in (B).

The observation of long-lasting ultrasound from liquid flow is not very surprising because in acoustic streaming, ultrasonic waves have to be continuously generated by a transducer to produce a steady-state flow. On the other hand, a continuous ultrasound does not necessarily guarantee a narrow directional flow. According to acoustic streaming, the streaming force is proportional to the ultrasound attenuation rate and in the same direction as ultrasound propagation (28–33), so a well-collimated flow must arise from a collimated ultrasound beam. Is our narrow stream driven by a directional ultrasound? The answer is also yes, and this is another big difference between ultrasounds generated by suspended Au particles and the Au particle–decorated cavity. On the basis of the pulse timing and speed of sound in water (1500 m/s), the first pulse in the inset of Fig. 4A comes from nanoparticles near the laser focus spot, and the second pulse is a reflection of the first pulse from the bottom of the cuvette (23). This understanding is illustrated in Fig. 4C, where we show ultrasound traces and their corresponding propagation pathways in three experimental configurations of a separate control experiment. When there is no streaming in case 1, ultrasound traces (black and red) consist of a pulse at 15 μs from suspended Au nanoparticles and its reflection at 32 μs. When streaming is created in case 2, the reflection in the ultrasound (blue) trace becomes much weaker and a strong pulse at 25 μs appears. This new pulse is the reflection from the back surface of the cuvette, and the delay of 10 μs from the first pulse is the time for the ultrasound to travel 1.5 cm from the front to back surface of the cuvette. Note that, in both cases, we have the Au nanoparticle solution, so the ultrasound has contributions from suspended Au particles as well as from the cavity. The simultaneous weakening of the reflection from the cuvette bottom and the emergence of reflection from the back surface are clear indications that the ultrasound from the cavity is not divergent as that from suspended Au nanoparticles, and it is propagating in the same direction as the transmitted laser. This directional propagation of ultrasound from the cavity is further confirmed by configuration 3, where the Au particle solution is replaced by pure DI water, and a glass is placed in the cuvette to reflect the ultrasound toward the hydrophone. As expected, a strong short pulse at 21 μs appears, and the pulse at 31 μs is a reflection from the side wall.

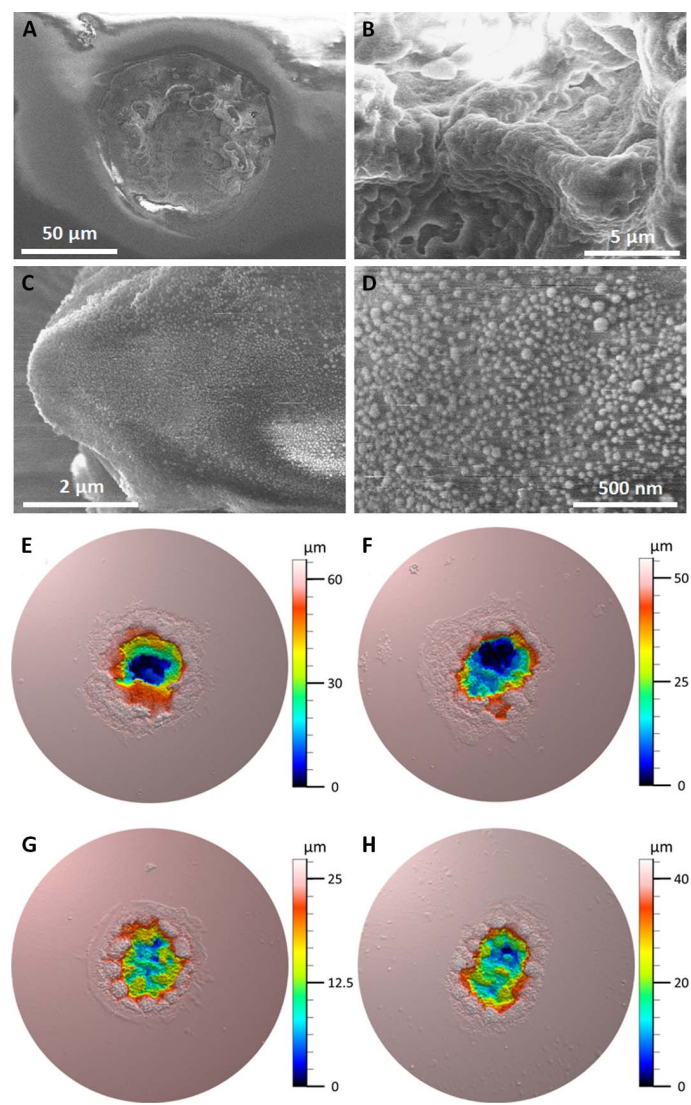

**Fig. 3. Electron micrographs and optical profiles of microcavities created by the laser.** (**A** to **D**) Scanning electron micrographs of a cavity and its surface created by the laser at normal incidence. (**E** to **H**) Optical tomographic images of cavities created by the laser at incident angles of (E) 0°, (F) 10°, (G) 20°, and (H) 30°. The diameter of each field of view is 200 μm.

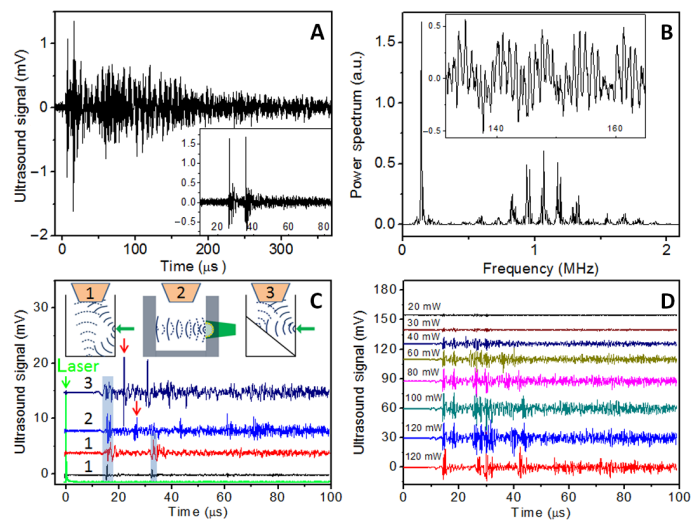

**Fig. 4. Ultrasound spectra and propagation pathways.** (**A**) Representative ultrasound traces (inset) without and with jets. (**B**) FFT spectrum of the ultrasound in (A). a.u., arbitrary units. Inset is the enlarged time domain traces from (A). (**C**) Ultrasounds in different configurations. Green, laser pulse. Ultrasound traces in (out of laser focus) black and (near focus) red are from configuration 1 without streaming. Trace in blue is from configuration 2 with streaming. Purple trace is from configuration 3 with streaming. (**D**) Laser power–dependent ultrasound signals. Trace in red uses configuration 2 in (C), and the rest use configuration 3 in (C). The size of the cuvette in (A) to (D) is 1.5 cm × 1.5 cm.

The established relationship between flow and ultrasound can help us understand the sudden disappearance of flow in Fig. 2 when power was reduced. Figure 4D shows a typical dependence of ultrasound on the laser power. The strength of ultrasound is not proportional to the laser power (34, 35). The decrease of ultrasound is very slow when the power is reduced from 120 to 80 mW. The transition happens around 60 mW, below which the ultrasound drops quickly, in agreement with the sudden weakening of the flow. As in configuration 3 of Fig. 4C, the insertion of a glass reflector changes the direction of both ultrasound and flow, but has no effect on their strength, indicating that the cavity in the front cuvette window determines the ultrasound and flow; a parallel back wall of the cuvette is not required. As can be seen in Fig. 4 (C and D), the periodic beating of ultrasound disappears after it is deflected by the inserted glass, and the flow does not have to be normal to the cuvette walls in Fig. 1. As to why long-lasting ultrasound can be generated by Au nanoparticles attached to the cavity, whereas suspended Au nanoparticles can only produce short pulse, we believe that this is due to a resonant mechanical coupling between the glass cavity and its attached nanoparticles (36).

Finally, let us try to understand how a plasmonic-acoustic cavity is created and how it is able to generate ultrasound and liquid stream that resembles the same laser that creates it. As a reminder, a suspension of Au nanoparticles is required to make such a cavity. Plasmonic Au nanoparticles have been shown to facilitate the creation of nanocavities or nanopits under pulsed laser irradiation in air (37, 38), but cavities with a size on the order of a micrometer or even tens of micrometers have not been reported. In general, we think that the formation of our plasmonic-acoustic cavity is a type of nanoparticle-assisted laser etching, but the underlying mechanism is different from mechanisms previously reported, namely, surface plasmon field–enhanced laser ablation and thermal melting/explosion of Au nanoparticles (37, 38). Both mechanisms will result in an extremely high local temperature. In our case, Au nanoparticles are surrounded by water instead of air, so thermal dissipation to and vaporization of the surrounding water are efficient especially for our 150-ns-long laser pulses (39, 40). Furthermore, nanoparticles attached to the cavity surface have sizes similar to the original ones, indicating that they are not fragments from the original nanoparticles. Note that ultrasound is also observed from suspended nanoparticles close to the laser spot on the cuvette wall; we think that some nanoparticles are ejected by ultrasound shock wave ultrasound against the glass wall, leading to the ablation of the glass and attachment of nanoparticles to the glass. Because the local shock wave depends on the laser intensity profile, the stronger the intensity is, the faster the sputtering is. As a result, the cavity reflects the laser profile, which depends on the incident angle. Because both processes involve ultrasound and nanoparticles, this could explain a kind of "reciprocal" relationship between laser and flow.

Flow images in Figs. 1 and 2 indicate that the streams have their maximum speeds very close to the cavities. The maximum speed of 4 cm/s with 120-mW laser is compatible with that observed in acoustic streaming driven by centimeter-sized ultrasound transducers (32, 41) and is 10 times faster than that needed for typical microfluidic devices (33). However, these streams have maximum speeds at a distance several times the diameters of transducers, which is the near-field boundary $L$ of the ultrasound, where $L = D^2/(4\lambda)$, $D$ is the source diameter, and $\lambda$ is the wavelength of the ultrasound (29, 30). In our case, $D$ is around 100 μm based on the flow images and actual cavity sizes from Fig. 3. $\lambda$ is 1.5 mm at the ultrasound center frequency of 1 MHz in Fig. 4. On the basis of this criterion, it is understood that the near-field region is very small in our case, but the same theory of ultrasound diffraction and acoustic streaming cannot explain the small divergence of our streams, which scales more like a laminar flow.

It is important to note that plasmonic nanoparticles are essential to the fabrication of the photoacoustic cavity and generation of ultrasound. The cavity and subsequent flow can be generated using 5-nm-diameter Au nanoparticles and even Au nanorods because both of them exhibit a strong plasmon resonance at the laser wavelength (figs. S1 and S2), but the laser streaming does not work for silver nanoparticles because their surface plasmon resonance is far away from 527 nm (fig. S3). A photoacoustic cavity is the most important component of the system because it connects photoacoustics to acoustic streaming. Certainly, the cavity's shape, size, and depth will be affected by many experimental parameters during fabrication such as the type and concentration of nanoparticles as well as the pulse width and focusing of the laser beam (figs. S4 and S5). Conversely, these dependencies give us more freedom and allow us to engineer the photoacoustic cavity to achieve the most efficient conversion. The real-time monitoring of cavity fabrication and flow also gives us an opportunity to obtain a nearly optimized cavity for the given laser focusing. Because photoacoustic streaming grows out from two well-known effects, its principle is general. Figure S6 shows that the same glass, when coated with ~50-nm Au film, can also generate a flow. However, the flow is much weaker and is always perpendicular to the glass surface regardless of the direction of the incident laser beam, which is consistent with the observation that the glass becomes covered by some Au particles after the ablation of the initial Au film. The cavity and flow can last for more than 1 hour (fig. S7). The aging of the cavity is believed to be due to the loss of Au nanoparticles, but this problem can be solved, for example, by using strongly attached nanoparticles and milder laser illumination.

## CONCLUSION

In conclusion, we have established a new optofluidic principle, demonstrated the generation of high-speed liquid flow using a pulsed laser, and developed a new technique of direct optical to mechanical transformation by combining the photoacoustic effect and acoustic streaming. The transformation is made through a plasmonic-acoustic cavity, which is completely self-fabricated and can generate a stream in the same direction as the laser. Although the basic principle behind the laser streaming is revealed, more research is needed to better understand the self-fabrication of the plasmonic-acoustic cavity, the resonant mechanical coupling between plasmonic nanoparticles and the cavity substrate, as well as the dynamics of the flow. Because the photoacoustic effect and acoustic streaming are well established, our laser streaming principle can be applied to any fluids and even gases (42). No interface or chemical modification of liquids is required because the driving force comes from the intrinsic acoustic dissipation in liquids at high frequency (16–20). Better photonic-acoustic transformers can be designed and fabricated for highly efficient optomechanical transformation, and the device can be made small or large to fit for different applications.

In addition to advancing this new interdisciplinary field involving photonics, acoustics, nanofabrication, fluid dynamics, and solid-state mechanics, laser streaming will find applications in optically controlled or activated devices such as microfluidics (32, 33, 41, 43), laser propulsion, laser surgery and cleaning, and mass transport or mixing.

### MATERIALS AND METHODS

The experimental setup is shown in Fig. 1 (A and B). A 527-nm pulsed laser beam (150-ns pulse width, 1-kHz repetition rate, average power < 130 mW) was focused with a 10-cm focal length lens on the wall of a glass cuvette. Gold nanoparticles (50 nm) were purchased from Ted Pella (part no. 15708-20). Fluorescent polymer microspheres (3.2 μm diameter) from Thermo Fisher Scientific were added to the water to visualize the flow. A PixeLINK color camera (PL-B742U) was used to capture the flow pattern with 50-ms exposure time.

### SUPPLEMENTARY MATERIALS

Supplementary material for this article is available at http://advances.sciencemag.org/cgi/content/full/3/9/e1700555/DC1

fig. S1. Comparison of flows generated by Au nanoparticles with different sizes.
fig. S2. Flows generated by Au nanorods.
fig. S3. No laser streaming with Ag nanoparticles.
fig. S4. Comparison of flows with different particle concentrations.
fig. S5. Streaming with different laser focus conditions.
fig. S6. Streaming with 50-nm Au film on glass in pure water.
fig. S7. Lifetime of streaming in pure water.
movie S1. Streaming and flow pattern.

**Acknowledgments**
**Funding:** J.B. acknowledges support from the NSF (CAREER Award ECCS-1240510) and the Robert A. Welch Foundation (E-1728). Z.W. is supported by the National Basic Research Program of China (973 Program, grant no. 2013CB933301) and National Natural Science Foundation of China (grant no. 51272038). S.D. acknowledges funding from the NSF